\documentclass[12pt]{iopart}

\usepackage{natbib}  
\usepackage{graphicx} 
\usepackage{rotating}
\usepackage{lscape}
\begin{document}

\title[Kabath et al. 2019]{Detection limits of exoplanetary atmospheres with 2-m class telescopes}

\author{P. Kab\'{a}th$^1$, Ji\v{r}\'{i} \v{Z}\'{a}k$^{2,1}$, H. M.J., Boffin$^3$, V. D. Ivanov$^3$, D. Jones$^{4,5}$, M. Skarka$^{2,1}$ }

\address{$^1$ Astronomical Institue, Czech Academy of Sciences, 
              Fri\v{c}ova 298, 25165, Ond\v{r}ejov, Czech Republic\\}
\address{$^2$ Department of Theoretical Physics and Astrophysics, Masaryk University, Kotl\'{a}\v{r}sk\'{a} 2, 60200 Brno, Czech Republic \\}
\address{$^3$ ESO, Karl-Schwarzschild-str. 2, 85748 Garching, Germany}
\address{$^4$ Instituto de Astrof\'isica de Canarias, E-38205 La Laguna, Tenerife, Spain Departamento de Astrof\'isica, Universidad de La Laguna, E-38206 La Laguna, Tenerife, Spain}
\address{$^5$ Universidad de La Laguna, E-38206 La Laguna, Tenerife, Spain}

\ead{petr.kabath@asu.cas.cz}     

\vspace{10pt}
\begin{indented}
\item[]March 2019
\end{indented}

\begin{abstract}

Transmission spectroscopy is an important technique to probe the 
atmospheres of exoplanets. With the advent of TESS and, in the future, of PLATO, more and more transiting planets around bright stars will be found and the observing time at large 
telescopes currently used to apply these techniques will not suffice. We demonstrate here that 2-m class telescopes equipped with spectrographs with high resolving power may be used for a certain number of potential targets. We obtained a time series of high-resolution FEROS spectra at the 2.2-m telescope at La Silla of the very hot Jupiter hosting planet WASP-18b and show that our upper limit is consistent with the expectations. This is the first analysis of its kind using 2-m class telescopes, and serves to highlight their potential. In this context, we then proceed to discuss the suitability of this class of telescopes for the upcoming flood of scientifically interesting targets from TESS space mission,
and propose a methodology to select the most promising targets. This is of particular significance given that observing time on 2-m class telescopes is more readily available than on large 8-m class facilities. 

\end{abstract}

%

\vspace{2pc}
\noindent{\it Keywords}: planetary systems ---  techniques: spectroscopic ---  planets and satellites: atmospheres --- 
 planets and satellites: individual: WASP-18

%
%
%

\section{Transmission Spectroscopy -- probe into the exo-atmospheres} \label{sec:intro}

The last decades have seen a wealth of discoveries of new exoplanets, mostly thanks to the transit detection method, either from the ground or from space.
The transit method not only allows us to detect candidate exoplanets, it also provides us with unique opportunities to probe
their physical properties, such as their bulk density and thus, indirectly, their composition. Moreover, transiting exoplanets offer the possibility to probe their atmospheric
composition, in particular through the technique of transmission spectroscopy \citep{TransSpec,dem2019}. This technique, which looks for the imprint a planetary atmosphere leaves
on the stellar light, is based on the fact that the atoms and molecules of the planetary atmosphere may block some fraction of the photons emitted by the host star. Thus, measuring the apparent radius of a planet as a function of wavelength informs us on the composition of the planetary atmosphere. This technique can be applied either using low- to mid-resolution, optimally time-resolved spectroscopy of the
star with a transiting planet, before, during and after the transit event \citep{SS00,ElyarNat}, or using high-resolution spectroscopy \citep{Snellen2008,Wytt15}, where spectra are obtained during and outside of the transit event, stacked and then compared to
one-another for detection of absorption in the planetary atmosphere during the primary transit. 

Because these techniques require data with very high signal-to-noise ratios -- the relevant effect being of the order of a few hundreds ppm at most --  such transmission spectroscopy is generally performed using instruments on 4-m or 8,10-m class telescopes (e.g. HARPS, CARMENES, FORS2), which are in very high demand. 
The detection and characterisation of exoplanetary atmospheres will become increasingly important with new targets which will be discovered soon by the TESS space mission \citep{TESS} and later by the  PLATO space mission \citep{PLATO}. TESS is expected to provide several thousands of new hot-Jupiter-sized exoplanets orbiting bright stars. Among these new planets, many systems will be good candidates for exo-atmosphere follow-up. It is clear that it will not be possible to follow-up all the interesting targets if we were to rely solely on the large telescopes that have been used up until now, especially as these will mostly be enlisted to study the most important (but also most challenging) targets. 

One of the first results on the detections of exo-atmospheres were obtained by Hubble-Space Telescope (HST) which is a 2-m class facility. However, the HST measurements were acquired with lower spectral resolving power \citep{wakeford,2018AJ....156..283E} allowing for resolution of absorption bands only. On the other hand high spectral resolving power provides a tool to obtain well resolved spectral lines which offer an insight into physical processes in the exo-atmospheres and probing high altitudes \citep{kempton}. However, it is much more challenging to obtain high signal-to-noise ratio (SNR) with high spectral resolving power. Therefore, it is interesting to assess whether ground-based 2-m class telescope equipped with spectrographs with high resolving power (40k and more) which can observe multiple transit events, can provide good results. 

It is also important to use as many different instruments as possible when doing transmission spectroscopy, given that the systematics of each instrument are still far from being understood. \cite{Gibson19}, for example, found that the features discovered by HST in WASP-31b could not be reproduced using either FORS2 or UVES, and they attribute this to a still incomplete understanding of the systematics of HST observations.

In this paper, we aim at verifying, using FEROS data, the suitability of 2-m class telescopes for performing transmission spectroscopy and constrain the parameter space in which such instruments can prove effective. Furthermore, we would like to stress again the advantage of the high resolving power and suggest that detection of absorbing features in exo-atmospheres is possible even with instrumentation on 2-m class ground-based telescopes.
The next section describes the FEROS data set. Results of our data analysis are described in the third section and detection limit, discussion and implications for planets expected to be found by TESS space mission are presented in section four.

\begin{table*}
\caption{Properties of the FEROS exposures. Individual RV errors for each point correspond to 0.040 km/s.}
\label{tab:spec}       
\footnotesize\rm
\begin{tabular}{lllllll}
\hline\noalign{\smallskip}
Exposure&ESO Data product file&Start Obs. (MJD)&End Obs. (MJD)&SNR&Airmass&RV (km/s)\\
\noalign{\smallskip}\hline\noalign{\smallskip}
1&ADP.2016-11-03T08:39:44.229.fits&57692.05330&57692.06372009&104.6&1.17&0.742\\
2&ADP.2016-11-03T08:39:44.237.fits&57692.06556&57692.07598116&124.7&1.14&0.682\\
3&ADP.2016-11-03T08:39:44.247.fits&57692.07781&57692.08823459&93.4&1.11&0.562\\
4&ADP.2016-11-03T08:39:44.255.fits&57692.09007&57692.10049253&112.5&1.09&0.402\\
5&ADP.2016-11-03T08:39:44.263.fits&57692.10234&57692.11276124&117.9&1.07&0.252\\
6&ADP.2016-11-03T08:39:44.269.fits&57692.11459&57692.12501270&118.4&1.06&0.042\\
7&ADP.2016-11-03T08:39:44.277.fits&57692.12686&57692.13727934&126.5&1.05&-0.128\\
8&ADP.2016-11-03T08:39:44.285.fits&57692.13913&57692.14954805&130.6&1.04&-0.258\\
9&ADP.2016-11-03T08:39:44.295.fits&57692.15138&57692.16179801&121.6&1.04&-0.378\\
10&ADP.2016-11-03T08:39:44.299.fits&57692.16363&57692.17405576&123.4&1.05&-0.478\\
11&ADP.2016-11-03T08:39:44.307.fits&57692.17591&57692.18632783&121.8&1.05&-0.598\\
13&ADP.2016-11-03T08:39:44.327.fits&57692.20044&57692.21085806&104.2&1.08&-0.838\\
\noalign{\smallskip}\hline
\end{tabular}
\end{table*}

\begin{figure}
\includegraphics[width=\hsize]{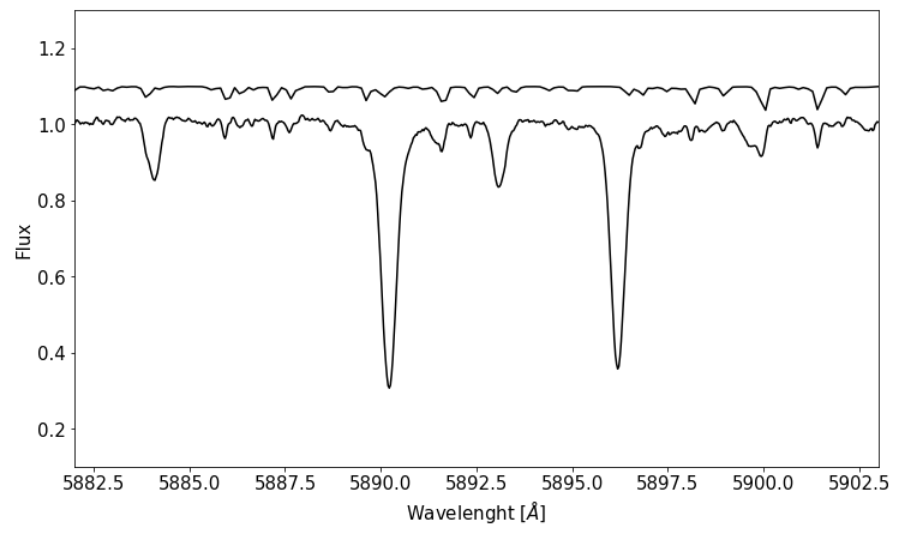}
\caption{The spectrum of WASP-18 in the vicinity of the sodium doublet, with a telluric model spectrum from {\tt Skycalc} plotted above.}
\label{fig:spec}       
\end{figure}

\section{Spectroscopic data} \label{sec:data}

We analyzed a time series of high-resolution spectra acquired using the FEROS instrument at 2.2-m telescope at La Silla, covering one transit of the hot Jupiter around its host star WASP-18. The data were obtained within the Czech observing slot during October 2016\footnote{TYCHO-BRAHE programme is supported by
the Ministry of Education, Youth and Sports project - LG14013} and the choice was very much limited as far as the target was concerned, as we had to find one  planet that was transiting during this period, visible for long enough from La Silla, and orbiting a bright enough star. The only object hosting a transiting planet that was available, WASP-18, has a $V$=9.3 magnitude. The 1.235$\pm$0.039 M$_\odot$, 1.215$\pm$0.048 R$_\odot$ F6 star is orbited by a very hot Jupiter ($T_{eq}=$ 2416 $\pm$ 58 K -- \cite{2019arXiv190107040G})  with an orbital period of 0.941 d. The planet itself has a mass of 10.43$\pm$0.30 M$_{\rm Jup}$ and a radius of 1.165$\pm$0.055 R$_{\rm Jup}$ \citep{W18}. In spite of its well-studied nature, this target and the acquired observations are perfectly suited to the purpose of this study - to probe the detection limits for 2-m class telescopes using real data with real systematics, and to ultimately constrain the role that such instrumentation can play in the studying the impending deluge of new transiting exoplanets.

The FEROS observations started on 31 October 2016 at 01:16 UT (MJD 57692.0533) and ended at 05:03 UT (MJD 57692.2109). The transit started at 01:44 UT and ended at 03:55 UT, so that we were able to obtain also a short out-of-transit phase. In total, 13 exposures of 900s each were obtained, 5 out of transit and 8 inside the transit. 

The FEROS spectrograph has a resolving power of 48,000 and covers the wavelength range from 360 to 920 nm \citep{kaufer95}. FEROS was used with one science fiber and another simultaneous ThAr fiber. For the analysis the pipeline-reduced data from the ESO Science Archive\footnote{http://archive.eso.org} were used. Each spectrum had a SNR between 93 and 131. The airmass at the start of the observations was 1.17 and it went down to 1.04, then up again to 1.08 at the end of the sequence. The seeing varied between 1" and 1.9" during the observations and the sky was clear. An example spectrum around the NaD region is presented in Fig.\ref{fig:spec}, while a summary of the observations is given in Table~\ref{tab:spec}. 

\begin{figure}
 \includegraphics[width=\hsize]{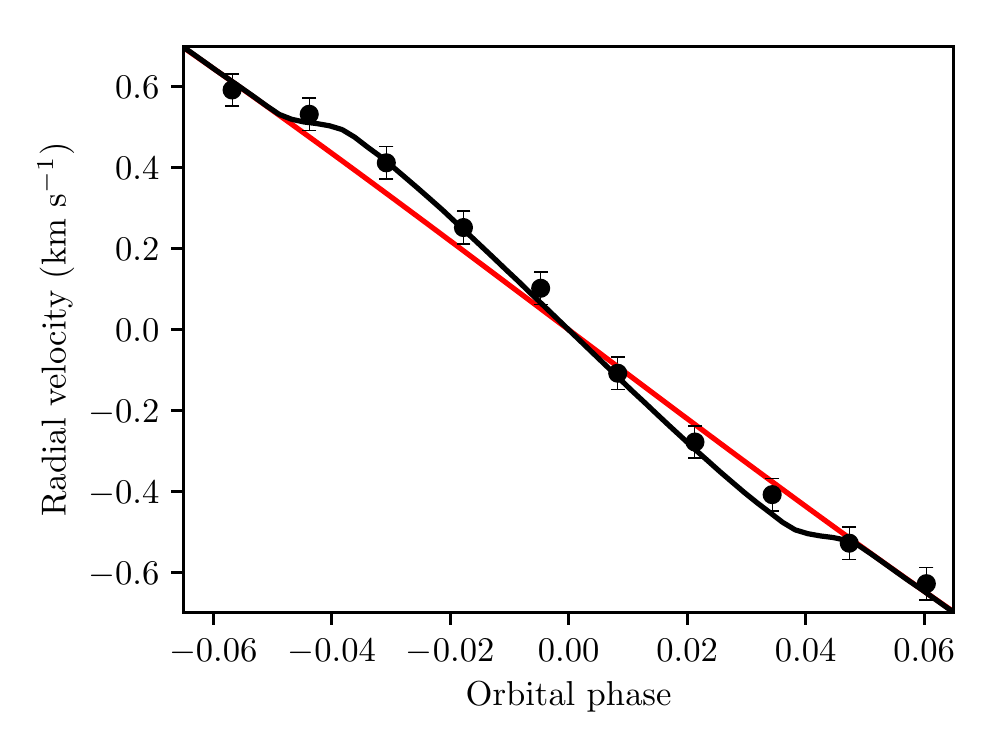}
\caption{FEROS radial velocities of WASP-18 as a function of the orbital phase. The lines overlaid show the flux-weighted (black) and dynamical (red) model radial velocities of the system, highlighting the clear detection of the Rossiter-McLaughlin effect.}
\label{fig:rv}       
\end{figure}

\section{Data analysis} \label{sec:analysis}

Our analysis follows the same procedure as that used by \cite{Wytt15} for the study of HD 189733. These authors analysed HARPS data covering 3 transits of this object to detect at a 10-$\sigma$ level the presence of sodium in the atmosphere of the planet. We reproduced their methodology on the same HARPS data, retrieved from the ESO Science Archive, to ensure that we could indeed obtain the same result. We then applied the same methodology to our set of FEROS observations. 

In brief, the FEROS pipeline reduced spectra\footnote{For detailed description of reduction process please see FEROS User Manual - http://www.eso.org/sci/facilities/lasilla/instruments/feros/doc/manual/P78/FEROSII-UserManual-78.0.pdf} with merged orders were used to determine the radial velocities from cross-correlation of our data with synthetic spectrum created by iSpec \citep{cuaresma14}. The calculated velocities are shown in Fig.~\ref{fig:rv} and in Table~\ref{tab:spec}. Each spectrum was corrected for the reflex motion of the star and as a reference a spectrum from mid-transit phase was used. We tried to apply the telluric correction as described in \cite{Wytt15}. However, due to only limited number of frames and a small airmass range, this method did not produce satisfactory results. We show a telluric spectrum obtained from ESO {\tt Skycalc}\footnote{https://www.eso.org/observing/etc/bin/gen/form? INS.MODE=swspectr+INS.NAME=SKYCALC} for comparison in Fig. 1 but we did not want to insert additional noise by subtracting the telluric lines. The telluric lines would be much more problematic for a region around the Potassium or TiO lines.

\subsection{Rossiter-McLaughlin effect}

Information about the Rossiter-McLaughlin (RM) effect can also be obtained from these observations. The RM effect can impact the detection limits of exo-atmospheres \citep{cegla} if not treated correctly or if it has a significant amplitude. In the case of WASP-18b this effect was already observed by \cite{tri} with the HARPS instrument. To demonstrate that we have similarly detected the effect with FEROS, we computed model radial velocities with the \textsc{phoebe} code version 2.1 \citep{prsa16,horvat18} using the system parameters derived by \cite{tri} and limb-darkening interpolated from model atmosphere tables within \textsc{phoebe} \cite[as described in][]{prsa16}.  Radial velocities were calculated for both the flux-weighted and dynamical (i.e.\ with and without the RM effect) regimes, and are shown overlaid on the observations in figure \ref{fig:rv}.  While the data do not have sufficient phase coverage to re-derive the system parameters, it is clear that the RM effect is detected, indicating that the performance of FEROS is, at least, suitable to exoplanetary radial velocity measurements and the detection of the RM effect.  As such, one can conclude that studies of the RM effect also represent an interesting science case for 2-m class telescopes.

\subsection{Exo-atmospheres}

The pipeline-reduced data were grouped into in- and out-of-transit spectra. Subsequently, a master out-of-transit spectrum was created by averaging all the out-of-transit spectra. Each in-transit spectrum was then divided by the master out-of-transit spectrum which was corrected for the computed radial velocity of the planet which is in the case of the WASP-18 system changing during the transit up to $\pm$ 60 km/s. For more details, see Eq. (3) of \cite{Wytt15}.

\begin{figure}
 \includegraphics[width=\hsize,angle=0]{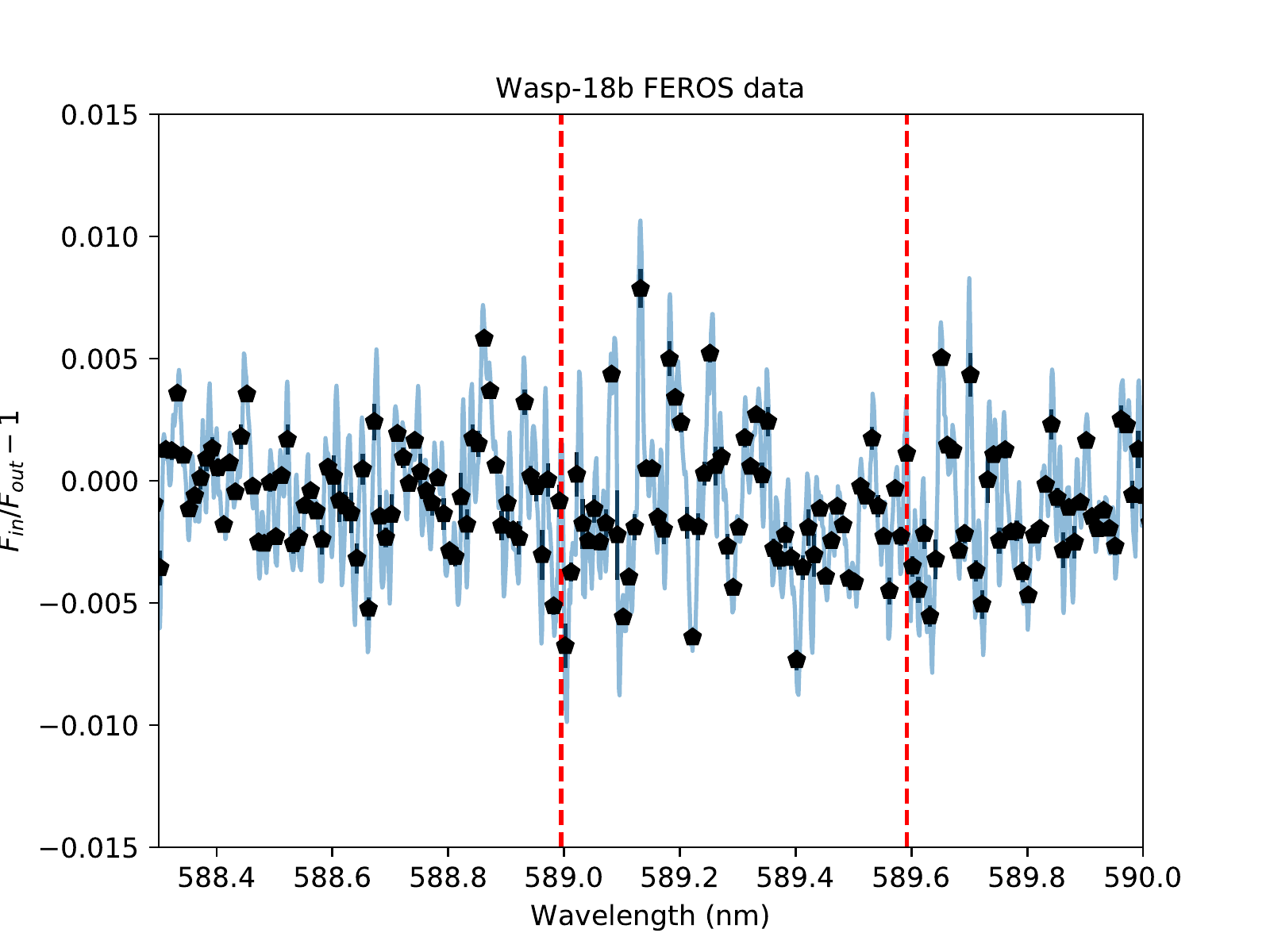}
\caption{The residuals of the ratio of the in- and out-of-transit mean spectra in the region of the sodium doublet (red lines). The data, in 10 point bins, are represented by the black dots.}
\label{fig:rms}       
\end{figure}

We selected several spectral regions of interest - around the sodium doublet, which is well-centered on order 26 of the FEROS spectra, and around H$\beta$, H$\alpha$. However, because of the negative results, we will only show here the region around the sodium doublet.
Figure~\ref{fig:rms} shows the ratio of in-transit averaged spectra over out-of-transit average spectra minus 1. As is obvious, we do not detect any sodium from the atmosphere of the planet. If we follow the procedure of \cite{Wytt15} and bin the data by 10 points\footnote{Wyttenbach et al. binned the data by 20 points in wavelength space. As FEROS' spectral resolution is about half that of HARPS, in order to be able to resolve at the same level the atomic lines, we only bin by 10 points here.}, we obtain a final r.m.s of 0.0022.

\cite{Wytt15} report with HARPS a 10-$\sigma$ detection for binned sodium lines which have 0.6\% depth, so their binned r.m.s is 0.0006, which is a factor 3.6 better than for the FEROS data. In any case, even using the same set-up and number of frames as they did, we wouldn't be able to detect any signal from the planet WASP-18b, given its too large gravity and therefore relatively small scale height. However, it is of greater interest to check whether the FEROS systematics are at the level we would expect if we would scale the HARPS detection limits. There are several factors that were in favour of the HARPS data: 
\begin{itemize}
\item the size of the telescope (3.6m compared to 2.2m, which corresponds to a gain of 2.67$\times$);
\item three transits were observed to accumulate more data;
\item WASP-18 is fainter than HD 189733 by 1.6 mag, which combined with the size of the telescope means that far longer exposures are needed for WASP-18 to reach a similar SNR.
\item the HARPS resolution is higher (115,000) than that of FEROS (48,000).
\end{itemize}
This allowed HARPS to obtain more data over the transit with similar signal-to-noise ratio (SNR). This is of course what ultimately needs to be compared. The HARPS data comprised 46 in transit spectra and 53 out of transit spectra, each of duration 300s or 600s, and having individual SNR between 45 and 150, with a combined SNR of 1026. 
For FEROS, we have only 13 exposures, for a combined SNR of 416, and we only bin by 10 points instead of 20. The expected difference in r.m.s. between the HARPS and FEROS data would thus be a factor of 3.5, which agrees very well with our measured value of 3.6.  

\begin{figure*}
 \includegraphics[scale=0.5,angle=0]{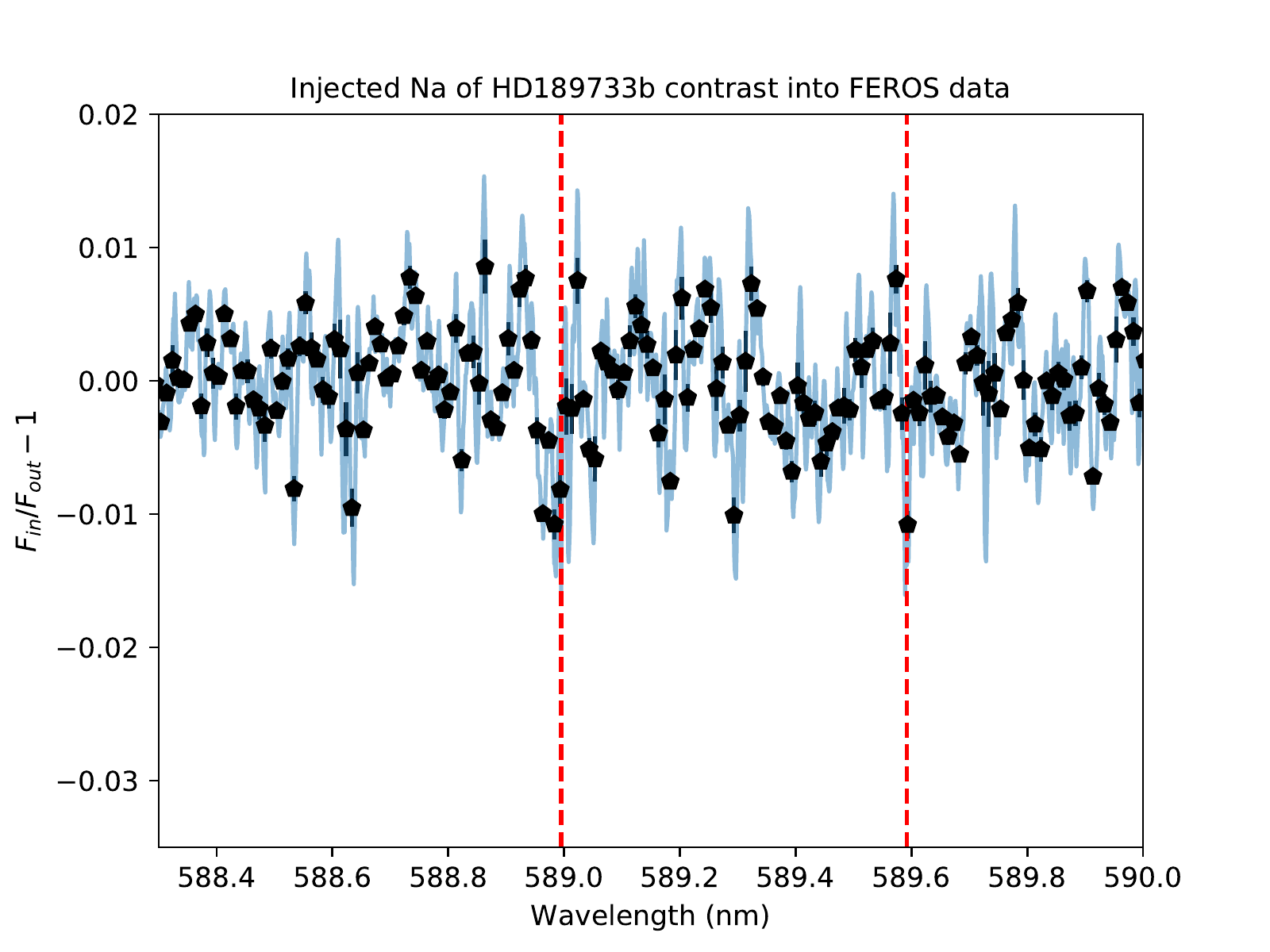}
 \includegraphics[scale=0.5,angle=0]{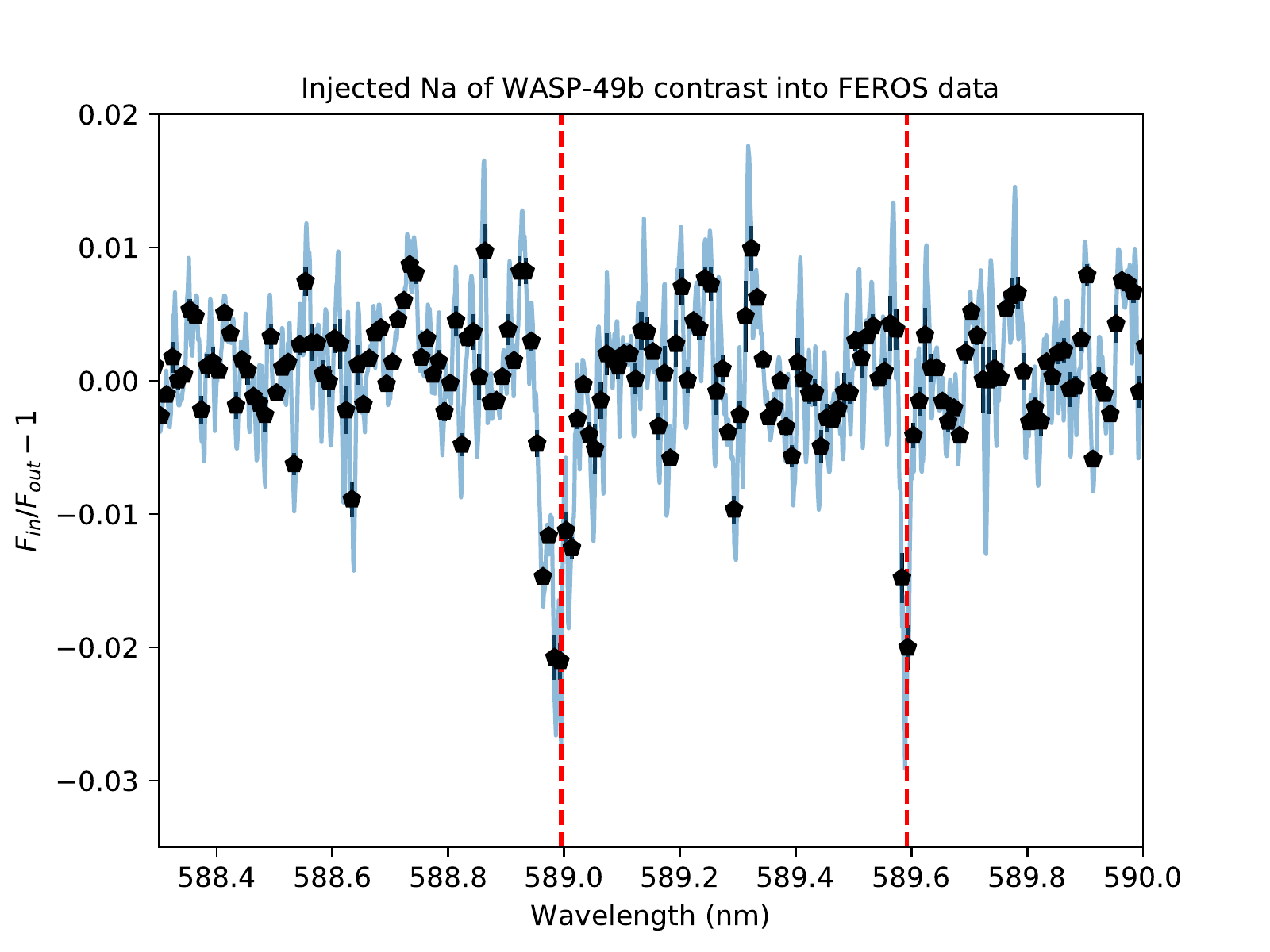}
\caption{Sanity test with injected signal: (left) the signature of sodium absorption in the planetary atmosphere was injected into the FEROS WASP-18b data set. The strength of the sodium signal was set to be equal to that detected from HD189733b by \cite{Wytt15}. (right) Injected sodium signal of the equivalent strength to the WASP-49b detection by \cite{wytt17}. Black points in both panels correspond to binning by a factor 10. The red dashed lines indicate the position of the NaD lines.}
\label{fig:NaDinj}       
\end{figure*}

As a final consistency check, an artificial sodium absorption signature from the planetary atmosphere was injected into FEROS WASP-18b data set. The injected signal corresponded to NaD with an abundance equal to that detected for HD 189733. The data were then re-analyzed as described above and the result is presented in Fig.\ref{fig:NaDinj}. In this test case, we detected the NaD signal with $2 \sigma$. This test proves that even for star with the brightness of WASP-18, we would be able to detect at least a lower limit of NaD in the exoplanetary atmosphere with a 2-m class telescope. We also show for comparison an injected signal of the equivalent strength of the WASP-49b detection reported by \cite{wytt17} in the same aforementioned Figure. In this case, the detection would be at the $4 \sigma$ level. We thus conclude that 2-m class telescopes can provide the necessary precision to detect absorption features in exoplanetary atmospheres if these are strong enough. 

In fact, had we observed HD 189733 (which has almost the same transit duration than WASP-18) with FEROS, we would inevitably have obtained a much larger SNR than that achieved for WASP-18b given the difference in brightness. But even with the current quality of data, our binned r.m.s. of 0.0022, would have allowed us to detect the sodium lines at a 2.7-$\sigma$, thus representing a lower limit. Furthermore, the detection level would have been further improved had we also observed three transits rather than just one. 
If one collects a sufficient number of exposures (either because the target is bright enough or by observing several transits), a 2-m class telescope would thus allow to do high-resolution transmission spectroscopy and detect signatures of atoms or molecules if they are as strong as those seen in HD 189733. There are increasing numbers of detections of sodium and other elements in exoplanetary atmospheres as reported by e.g.\cite{wytt17,rojo13} - with both discoveries made by teams using 3.6-m or larger aperture telescopes.  As such, we now turn to the question of how many planets might represent feasible targets for smaller aperture facilities. 

\begin{figure*}
  \includegraphics[width=\textwidth,angle=0]{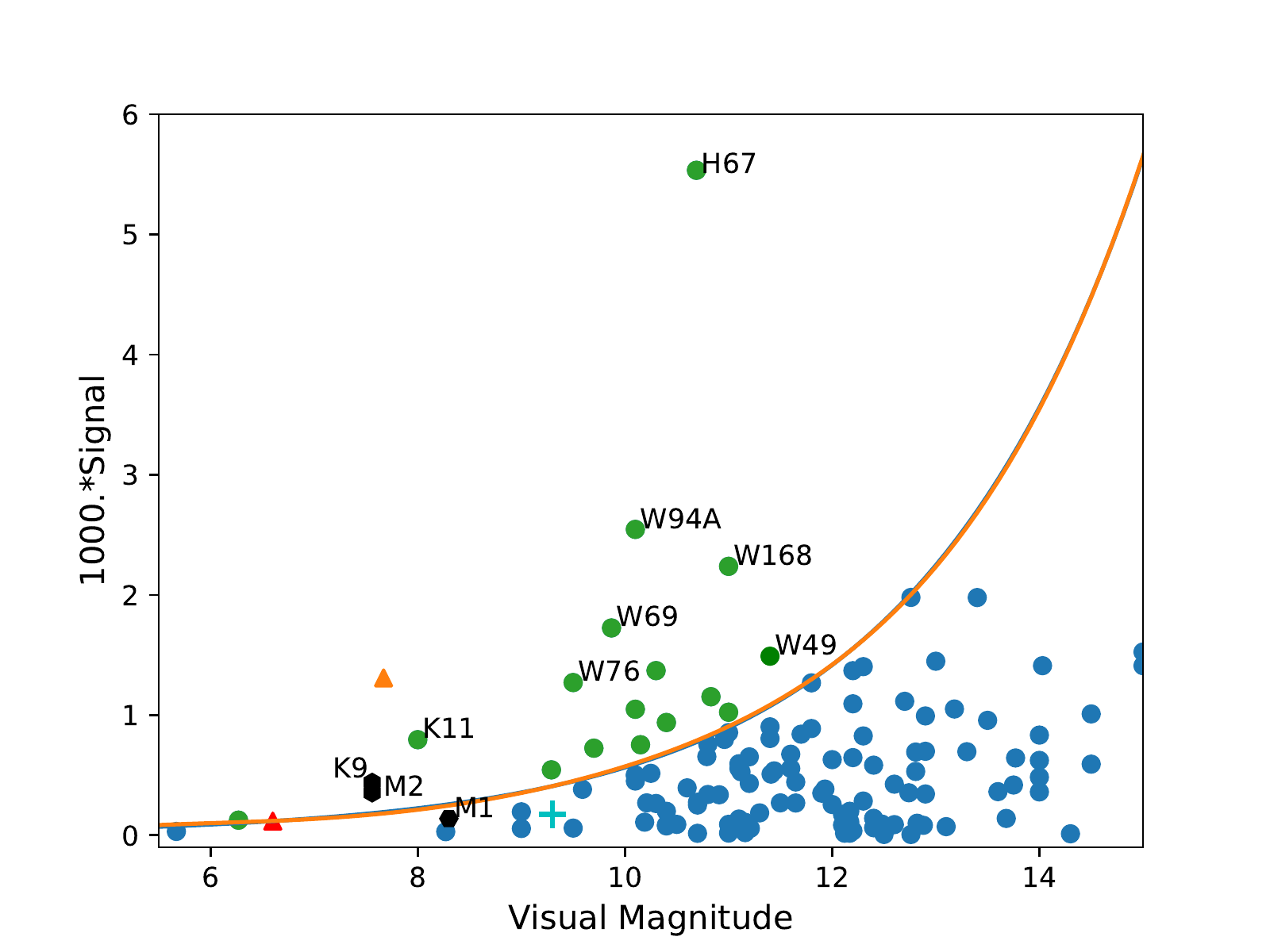}
  
\caption{The expected atmospheric signal for well-characterised transiting planets as a function of the visual magnitude of the host star. The orange triangle shows the position of HD 189733, while the red triangle is the rescaled value, assuming a 2m-class telescope (see text). All the points above the solid line are suitable candidates to perform transmission spectroscopy with a 2m-class telescope. 
}
\label{fig:diag}       
\end{figure*}

\begin{figure*}
   \includegraphics[width=\textwidth,angle=0]{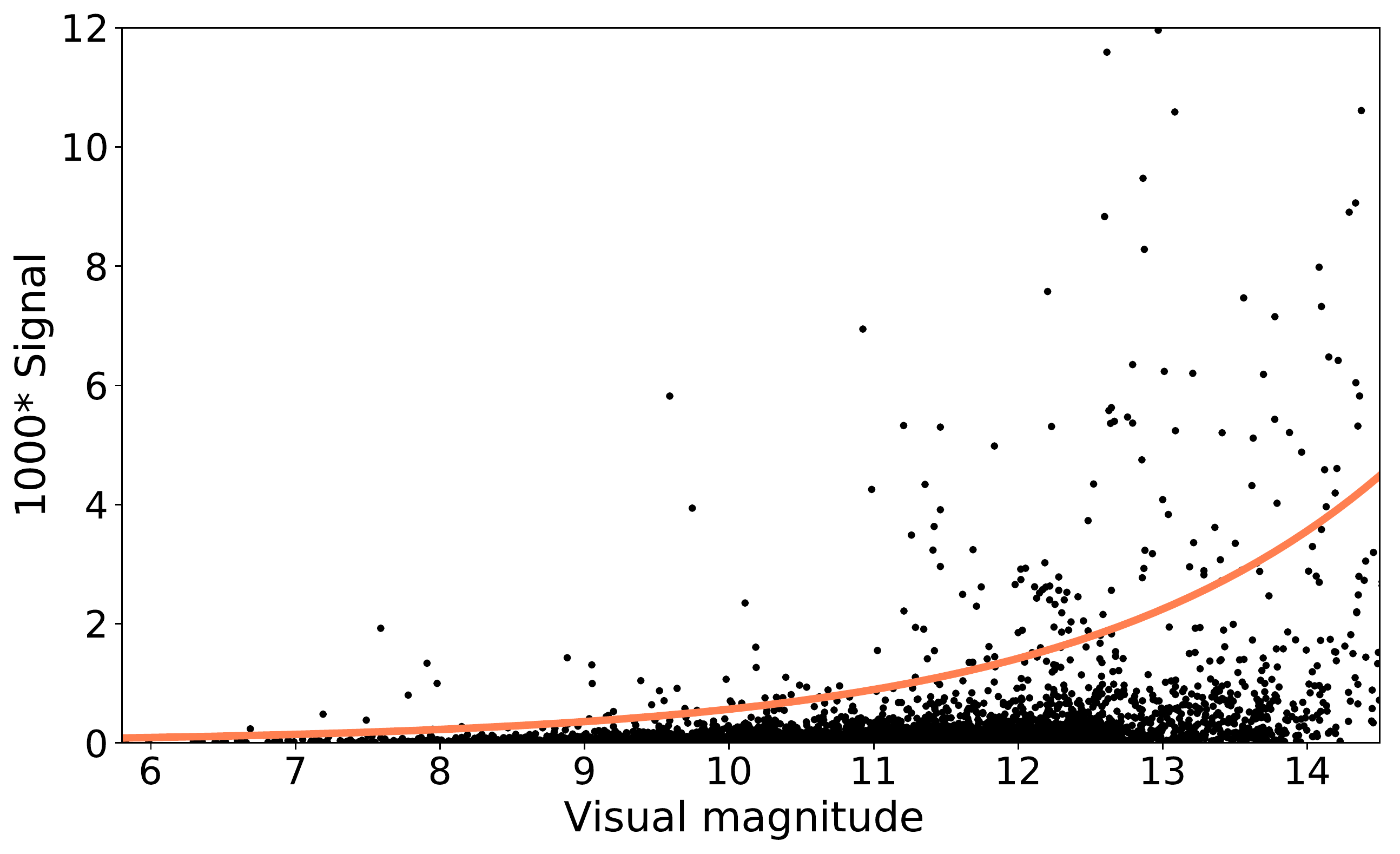}
\caption{Expected signal as in previous Fig. \ref{fig:diag} but from TESS planets which are depicted as black dots. The orange line is the detection limit for 2-m telescopes and good candidates for further follow-up will be above the line (see text for more detail).}
\label{fig:tess1}       
\end{figure*}

\section{Detection limits of 2-m class telescopes and implication for TESS era} \label{sec:limits}

The signature of a planetary atmosphere, $\Delta \delta$, is roughly given by 
\begin{equation}\label{eq:sig}
\Delta \delta \simeq \frac{2 n H R_p}{R_s^2},
\end{equation}
with $R_s$, the stellar radius, and $n$, the number of atmospheric scale heights ($H$), to consider, which is given by 
\begin{equation}
H = \frac{k_B T_p}{\mu_m g},
\end{equation}
where $k_B$ is the Boltzmann constant, $T_p$ is the temperature of the planetary atmosphere, $\mu_m$ is the atmospheric mean molecular weight and $g$ is the planet surface gravity, $g=G M_p/R_p^2$, with $G$, the gravitational constant, and $M_p$ and $R_p$ being the mass and radius of the planet, respectively. In the following, we will assume $\mu_m = 2.2$, and take $n=6$ corresponding to high altitudes. We choose, $n=6$ because sodium absorption was reported in high altitudes by e.g. \cite{2008ApJ...673L..87R} and \cite{Snellen2008}, Similarly a high altitude hydrogen absorption was reported by \cite{2018A&A...616A.151C}. However, we recognise, that the $n=6$ assumption is valid mostly for strong absorbers and with decreasing n, the corresponding signal will be reduced. As an example, the TiO features in Wasp-19b were reported for $n \sim 2$ \citep{ElyarNat}. When no measured value of $T_p$ is known, we will use the equilibrium temperature, which depends on the albedo of the planet, the temperature and radius of the star, as well as the separation between the star and the planet.

We downloaded from the Extrasolar Planets Encyclopaedia\footnote{exoplanet.eu} all the transiting planets for there was enough information to compute the expected signal, as given by Eq.~\ref{eq:sig}. A plot of all these planets is shown in Fig.~\ref{fig:diag}. We show there the position of HD 189733 as an orange triangle. We then rescale this point, such that we only require a 3-$\sigma$ detection (instead of the 10-$\sigma$ detection by HARPS) and assuming we have a 2m telescope instead of the 3.6m on which HARPS is mounted. The latter means that if we want to obtain also 99 spectra (as was the case for HARPS) at the same SNR, we would need to observe an object that is brighter. Alternatively, this would also be achieved if the transit duration is longer than the 2 hours of HD 189733 or if we would observe more than 3 transits. This may be an issue, however, if the host star is quite active or if the spectrograph is not stable for a long period of time, as observing more than 3 transits would possibly require observing over several months, or even different seasons. The rescaled point is shown as the red triangle in the plot. We can then estimate what is the amplitude of the signal that is needed to observe a star of a given magnitude -- this is shown by the blue curve. Thus, any object that is above that line (shown as green dots) \emph{that would have atomic or molecular signature at the same level as HD 189733} could, in principle, be studied successfully with a 2m-class telescope. We note, however, that in practice, one would be limited to bright stars given that at, say, $V=10$\,mag an exposure of roughly 30 min with FEROS on the 2.2m telescope would be required. This would be, in most cases, the longest exposure time that could be consider before smearing out the planetary signal and/or failing to acquire sufficient spectra during each individual transit. As we have now TESS mission detecting the first planets, we exploited TESS data published by \cite{2018ApJS..239....2B} who presents
stellar and planetary parameters of TESS targets. We employed mass-
radius relation from \cite{2017A&A...604A..83B}. Equilibrium temperature of 
the planet was calculated using Eq. 1 from \cite{2007ApJ...669.1279S}. Calculated
data are shown in Fig.~\ref{fig:tess1}. Furthermore, all suitable candidates for exo-atmosphere characterization with
sufficient SNR were checked with Gaia DR2 stellar radius values to exclude 
the possibility of red giant contamination. It can be seen, that TESS should provide several dozens of good candidates (expected to be in similar regions as known good candidates in Fig \ref{fig:diag}) for characterization of exo-atmospheres, even for 2-m class telescopes.

Fig.~\ref{fig:diag} shows that among the known transiting planets, there are many candidates that could be studied with 2-m class telescopes using transmission spectroscopy. This, of course, assumes that these objects have planetary signatures at the same level as HD 189733. This may not be the case, for example if there are clouds in the atmosphere. However, it still shows us that in this case, a non-detection would still be significant. 

We highlight on the plot some specific examples (in black). Among the brightest objects, we indicated the positions of KELT-9 (K9), MASCARA-1 (M1) and MASCARA-2 (M2). While it appears clear that MASCARA-1 is out of reach, KELT-9 and MASCARA-2 (which are basically at the same position in the diagram) are potential targets, in fact also known to present clear atomic or molecular signatures \cite[e.g.][]{Mascara2}. Similarly, among those stars with the highest signal, let us single out KELT-11 (K11), WASP-49 (W49), WASP-69 (W69), WASP-76 (W76), WASP-94 A (W94A), WASP-168 (W168), and HAT-P-67 (H67), as indicated in the figure. For the latter, the expected signal is very strong and given that it has a transit duration of 4.7 hrs, it may still be a prime candidate for 2m-class telescopes, despite its magnitude of 10.7.
WASP-18 itself is shown as the cyan cross in the diagram and it is obvious from its position that it would not have been possible to detect it with FEROS\footnote{As noted above, unless of course, we obtain many more exposures during several transits.}. 

\begin{figure*}
  \includegraphics[width=\textwidth,angle=0]{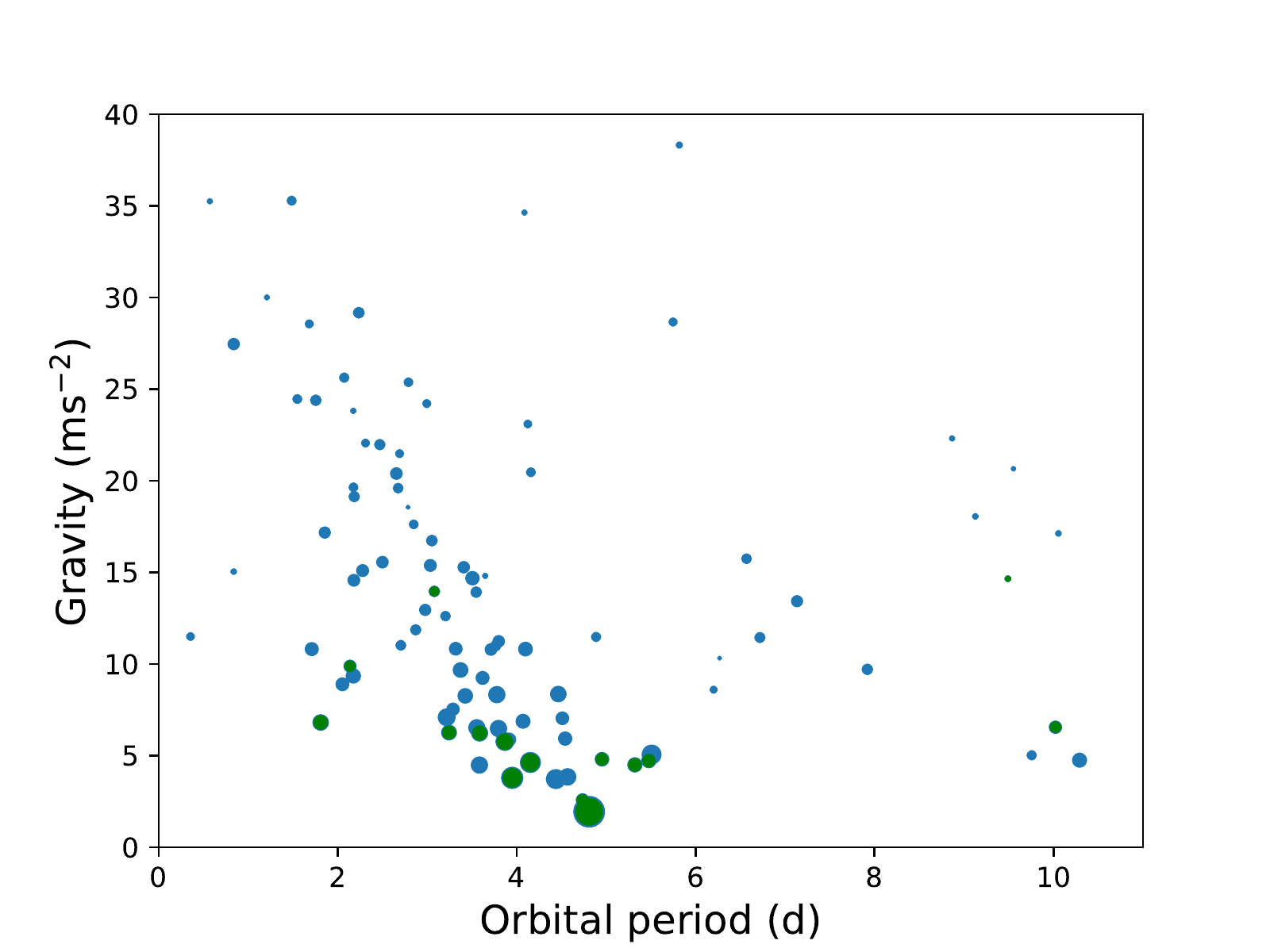}
\caption{The orbital period - planetary gravity plane showing all the targets from Fig.~\ref{fig:diag} with the size of the dots proportional to the expected atmospheric signal. The objects in green are those that are feasible with 2m-class telescopes.}
\label{fig:sig}       
\end{figure*}

\begin{figure*}
  \includegraphics[width=\textwidth,angle=0]{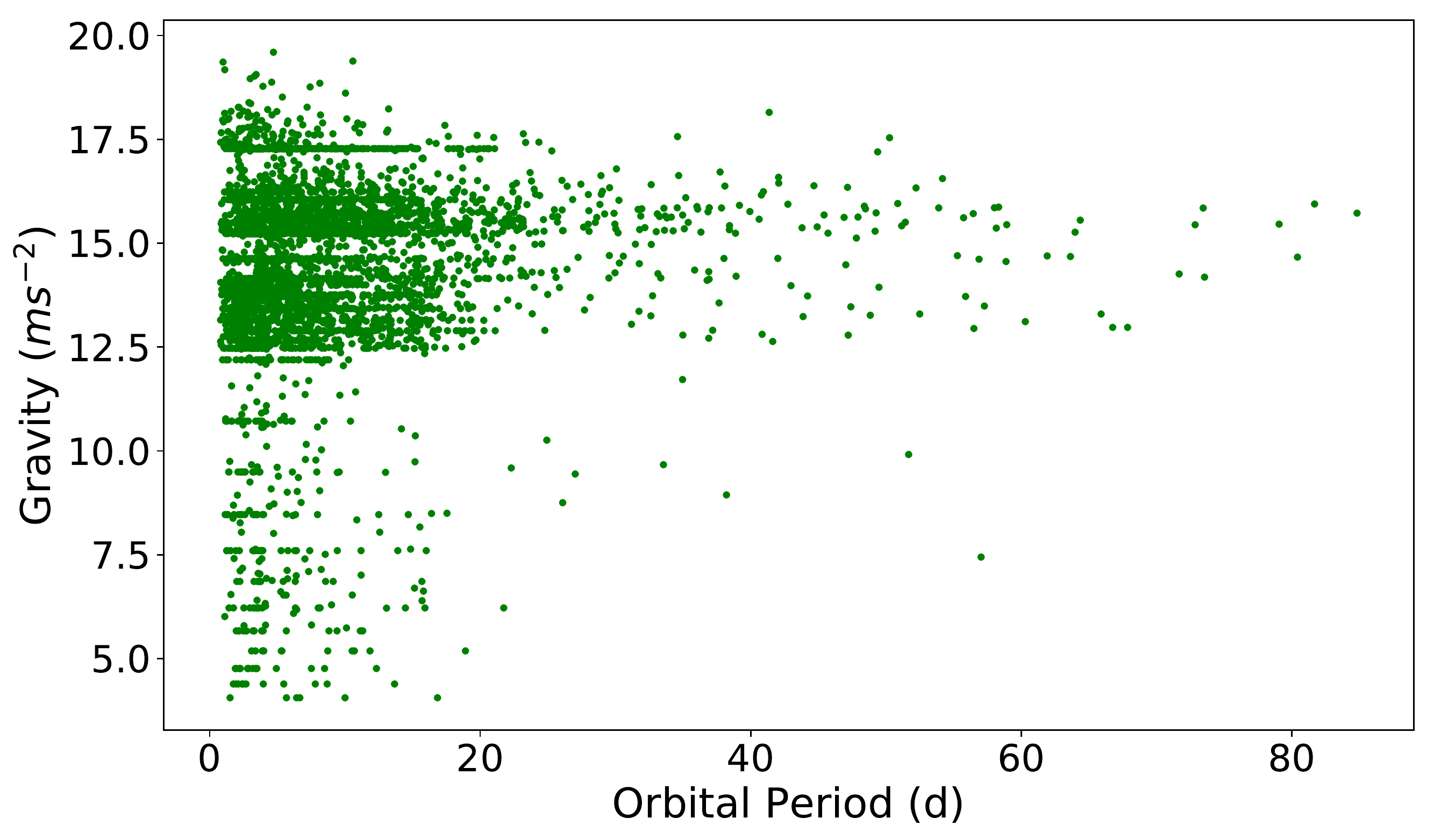}
\caption{Green points represent the expected distribution of gravity of TESS planet versus their orbital periods. Most of the good candidates for follow-up observations will be in lower left corner below 10 ms$^{-2}$.}
\label{fig:tess2}      
\end{figure*}

Eq.~\ref{eq:sig} shows that large planetary atmosphere scale height and a smaller host
star increase the signal. The first requirement implies a low planetary gravity \footnote{This is why WASP-18 is a very bad target as it has a very large gravity.} and a high temperature. The temperature of the planet is directly related to the host star's temperature and its distance from the host, or equivalently, the orbital period. When looking at solar-like stars -- i.e., assuming that the ranges of stellar masses, radii and temperatures are limited --  the main effect is thus due to the gravity of the planet and, to a lesser extent, to the orbital period. This is visible in Fig.~\ref{fig:sig}, where it is obvious that the most promising targets are essentially constrained in the region ($ g < 10~{\rm ms^{-2}}, P < 6$ d). We expect the similar situation for TESS planets presented in Fig. ~\ref{fig:tess2} which also shows many potentially good candidates.

\begin{landscape}
\begin{table}

\caption{\label{tab:cand}Planets for which the atmospheric signal may be detected.}

\footnotesize\rm

\begin{tabular*}{1.5\textwidth}{lcccccccccccccc} 
\br
Name & RA DEC & $V$ & $M_p$ & $R_p$ & $T_p$ & $M_s$ & $R_s$ & Sp. & T$_{\rm eff}$ & g &  H& $\Delta \delta$ & Reference\\
 & 
& 
mag & 
M$_{\rm Jup}$ & 
R$_{\rm Jup}$ &
K & 
M$_\odot$ & 
R$_\odot$ &
Type & 
K &
ms$^{-2}$ &  
km  & 
 & \\

\br

MASCARA-2~~ &19:38:38.7 $+$31:13:09& 7.6&3.52  &1.74&2230&1.89 & 1.60 & A2 & 8980 & 28.82 &292 &0.34&  \cite{Mascara2}\\
KELT-9      &20:31:26.4 $+$39:56:20& 7.6&2.88&1.89&4050&2.52 & 2.36 & A0 & 5370 &  20 & 764& 0.44& \cite{gaudi17} \\
KELT-11     &10:46:50.0 $-$09:23:57& 8.0&0.19&1.37&1712&1.43&2.72&G8&5370&2.58&2502&0.79 & \cite{pepper17}\\
KELT-17     &08:22:28.0 $+$13:44:07& 9.3&1.31&1.52&2087&1.63&1.64&A&7454&13.96&564&0.54& \cite{zhou16}\\
WASP-76     &01:46:32.0 $+$02:42:02& 9.5&0.92&1.83&2190&1.46&1.73&F7&6250&6.81&1212&1.27& \cite{west16} \\
WASP-74     &20:18:10.0 $-$01:04:33& 9.7&0.97&1.56&1910&1.48&1.64&F9&5970&9.88&729&0.72&\cite{hellier15} \\
WASP-69     &21:00:06.0 $-$05:05:40& 9.9&0.26&1.05&963 &0.82&0.81&K5&4715&5.77&629&1.72 & \cite{anderson14}\\
WASP-131    &14:00:46.0 $-$30:35:01&10.1&0.27&1.22&1400&1.06&1.53&G0&6030&4.50&1173&1.05 & \cite{hellier17}\\
WASP-94A    &20:55:07.9 $-$34:08:08&10.1&0.45&1.72&1604&1.29&1.36&F8&6170&3.79&1596&2.54& \cite{malle14}\\
WASP-117    &02:27:06.0 $-$50:17:04&10.2&0.27&1.02&1024&1.12&1.17&F9&6040&6.55&589&0.75& \cite{lendl14}\\
WASP-101    &06:33:24.0 $-$23:29:10&10.3&0.50&1.41&1560&1.34&1.29&F6&6400&6.23&944&1.37&\cite{hellier14b}\\
WASP-88     &20:38:03.0 $-$47:32:17&10.4&0.56&1.70&1775&1.45&2.08&F6&6431&4.80&1394&0.94&\cite{delrez14}\\
HATP-67     &17:06:27.0 $+$44:46:37&10.7&0.34&2.08&1903&1.64&1.54&F&6406&1.94&3698&5.53& \cite{zhou17}\\
KELT-8      &18:53:13.0 $+$24:07:39&10.8&0.87&1.86&1675&1.21&1.67&G2V&5754&6.26&1009&1.15&\cite{fulton15}\\
WASP-168    &06:26:59.0 $-$46:49:17&11.0&0.42&1.50&1340&1.08&1.12&F9V&6000&4.63&1091&2.24 &\cite{hellier19}\\
WASP-172    &18:53:13.0 $+$24:07:39&11.0&0.47&1.57&1740&1.49&1.91&F1V&6900&4.73&1387&1.02&\cite{hellier19}\\
WASP-49     &06:04:21.5 $-$16:57:55&11.4&0.39&1.19&1400&1.00&1.03&G6V&5600&6.83 &773 & 1.49 &\cite{wytt17}\\

\br
\end{tabular*}

\end{table}
\end{landscape}

Although we have seen that there are already targets that could be successfully observed with 2m-class telescopes, the situation will improve thanks to the targets that will be found by TESS and by PLATO.
Of the 471 candidate transiting planets found by TESS as of March 2019, about a half of them had a magnitude below or equal to 10.5 and thus are ideally suited for 2-m class telescopes. 
Moreover, 41 of these have a planetary radius larger than 1 R$_{\rm Jup}$ and an orbital period below 6 days. These are the conditions such that the signal should be detectable. If we assume that the candidates
discovered until now are representative of the whole mission, we will thus have potentially 22\% of all TESS candidates that could be useful targets for transmission spectroscopy with 2-m class telescopes. The same
telescopes could, of course, also first be used to confirm the planetary status of these candidates, determine the orbits and the planetary masses, and therefore allow a finer selection of suitable targets for transmission spectroscopy. Furthermore, it is important to stress that 2-m class telescopes are suitable for long-term monitoring or for multiple transit events observations. The effects of stellar variability which might influence multiple nights approach can be well monitored with Ca I (612.2217 nm and 616.2173 nm) and Mg I lines (518.3604 nm), as described in \cite{wytt17}. We have demonstrated that multiple transit events are almost a must for a successful detection of transmission features in exo-atmospheres, and given that it is intrinsically easier to obtain this telescope time on 2-m class facilities it makes their use all the more attractive. 
Given that TESS is expected to have at least several thousands of candidates at the end of its mission, the future of 2-m class telescopes is very bright indeed.

\section{Summary} \label{sec:conc}

 It is not straightforward to assume that exo-atmospheres can be probed with smaller class telescopes. The brightness of the target is not the only factor which needs to be considered. Even for some 4-meter class telescopes this task might not be feasible when not performed using appropriate instrumentation and methodology. We provide here also clear guidance for feasibility studies and for Telescope Time Allocation committees which are are of broad interest for the community. There are many under-employed, and sometimes even neglected, 2-m class telescopes with reasonable instrumentation which could dedicate significant amounts of time to the study of exo-atmospheres in the era of TESS and PLATO. Especially, spectrographs with high resolving power can be very suitable for such studies because compared to spectrophotometry with low resolving power, they are not limited by the comparison stars. In this paper, we have shown for the first time that it is reasonable to perform transmission spectroscopy of bright enough targets with such facilities, provided a few rules are followed:
\begin{itemize}
\item For a given planet which is bright enough (typically $V<10$), compute the expected signal based on Eq.~\ref{eq:sig}, and using its $V$ magnitude verify that it is above the threshold given by (orange line in Fig. \ref{fig:diag} )
$$\Delta \delta = 1.179 \times 10^{-4+0.2  (V-6.6)} $$.
\item  Take into account the expected number of scale heights $n$ (here n=6 for high altitude Na as in HD209458b). However strong features like sodium were proven in high altitudes. Spectroscopy with high resolving power is sensitive for probing of high altitudes. 
\item Make sure that the exposure time to obtain spectra with a reasonable SNR ($> 50$)  is short enough to obtain at least 5  spectra during one transit providing a reasonable time sampling
\item Obtain enough out-of-transit and in-transit spectra (ideally equal numbers) so that the total SNR is larger than 500. Observations over several transits may be necessary.
\item Stellar activity can be monitored on Ca I and Mg I lines  
\end{itemize}

We have also provided ideal candidates for which such studies could be performed and have shown that TESS should provide a flurry of further targets accessible with 2-m class telescopes.

\ack
We would like to thank to the anonymous referee for useful comments. The data presented here were obtained via the TYCHO-BRAHE programme supported by
the Ministry of Education, Youth and Sports project - LG14013, and are available in the ESO archive under ESO progamme ID 098.A-9039(C) which was taken with Max Planck 2.2-m telescope with FEROS. 
PK and JZ would like to acknowledge the support from GACR international grant 17-01752J. MS acknowledges financial support of Postdoc@MUNI project CZ$.02.2.69/0.0/0.0/16\_027/ 0008360$. This work benefited from support for mobility by the DAAD-18-08 programme, Project-ID: 57388159. DJ acknowledges support from the Spanish Ministry of Economy and Competitiveness (MINECO) under the grant AYA2017-83383-P. We would like
to thank the observer Adela Kawka. We acknowledge the use of TESS Alert data which is currently in a beta test phase, from pipelines at the TESS Science Office and at the TESS Science Processing Operations Center. We acknowledge the usage of following software - astropy \cite{2013A&A...558A..33A,2018AJ....156..123A} and Phoebe \cite{prsa16,horvat18},
iSpec \cite{cuaresma14}

\newcommand{\newblock}{}

\end{document}